\documentclass{entcs}
\usepackage{prentcsmacro}

\newif\ifpdf
\ifx\pdfoutput\undefined
  \pdffalse
\else
  \pdfoutput=1
  \pdftrue
\fi

\ifpdf
  \usepackage[pdftex]{color}
  \usepackage[pdftex]{graphicx}
  \DeclareGraphicsExtensions{.pdf}
\else
  \usepackage[dvips]{color}
  \usepackage[dvips]{graphicx}
  \DeclareGraphicsExtensions{.eps}
\fi

\usepackage{times} 
\usepackage{code}
\usepackage{latexsym}
\usepackage{amsmath}
\newcommand{\secref}[1]{Section~\ref{#1}}
\newcommand{\tblref}[1]{Table~\ref{#1}}
\newcommand{\figref}[1]{Figure~\ref{#1}}

\newcommand{\eg}{{\em e.g.}}

\newcommand{\ie}{{\em i.e.}}

\newcommand{\role}{r\^{o}le}

\newcommand{\C}{\textsc{C}}
\newcommand{\java}{\textsc{Java}}

\newcommand{\moby}{\textsc{Moby}}

\newcommand{\ml}{\textsc{ML}}
\newcommand{\sml}{\textsc{SML}}
\newcommand{\smlnj}{\textsc{SML/NJ}}

\newcommand{\ocaml}{\textsc{OCaml}}

\newcommand{\haskell}{\textsc{Haskell}}

\providecommand{\bftt}[1]{{\ttfamily\bfseries{}#1}}
\providecommand{\ittt}[1]{{\ttfamily\itshape{}#1}}
\providecommand{\kw}[1]{\bftt{#1}}

\newcount\timeHH
\newcount\timeMM
\timeHH=\time
\divide\timeHH by 60
\timeMM=\time
\multiply\count255 by -60 \advance\timeMM by \count255
\newcommand{\timestamp}{%
  \today{} --- 
  \ifnum\timeHH<10 0\fi\number\timeHH\,:\,\ifnum\timeMM<10 0\fi\number\timeMM}

\newcommand{\TOOL}[1]{\textit{#1}}
\newcommand{\cut}[1]{}

\newcommand{\bol}{BOL}
\newcommand{\mbi}{MBI}

\newcommand{\charon}{\TOOL{Charon}}
\newcommand{\mobyidl}{\TOOL{moby-idl}}

\renewcommand{\codedisplaysize}{\footnotesize}

\SetMathAlphabet{\mathtt}{normal}{OT1}{pcr}{n}{n}
\SetMathAlphabet{\mathtt}{bold}{OT1}{pcr}{bx}{n}

\def\lastname{Fisher, Pucella, and Reppy}

\begin{document}

\begin{frontmatter}
\title{
  A framework for interoperability
}
\author{Kathleen Fisher\thanksref{ksf-email}}
\address{
  AT\&T Labs, Research \\
  180 Park Ave., E244 \\
  Florham Park, NJ 08932}
\author{Riccardo Pucella\thanksref{riccardo-email}}
\address{
  Department of Computer Science \\
  Cornell University \\
  Ithaca, NY 14853}
\author{John Reppy\thanksref{jhr-email}}
\address{
  Bell Labs, Lucent Technologies \\
  600 Mountain Ave. \\
  Murray Hill, NJ 07974}
\thanks[ksf-email]{
  Email: \texttt{kfisher@research.att.com}}
\thanks[riccardo-email]{
  Email: \texttt{riccardo@cs.cornell.edu}}
\thanks[jhr-email]{
  Email: \texttt{jhr@research.bell-labs.com}}
 
\begin{abstract}
Practical implementations of high-level languages must provide
access to libraries and system services that have APIs specified
in a low-level language (usually \C{}).
An important characteristic of such mechanisms is the
\emph{foreign-interface policy} that defines how to bridge the semantic gap
between the high-level language and \C{}.
For example, IDL-based tools generate code to marshal data into and out of
the high-level representation according to user annotations.
The design space of foreign-interface policies is large and there are
pros and cons to each approach.
Rather than commit to a particular policy, we choose to focus on
the problem of supporting a gamut of interoperability policies.
In this paper, we describe a framework for language interoperability that 
is expressive enough to support very efficient
implementations of a wide range of different foreign-interface policies.
We describe two tools that implement substantially different policies
on top of our framework and present benchmarks that demonstrate their
efficiency.
\end{abstract}%

\end{frontmatter}

\section{Introduction}
\label{sec:intro}
High-level languages, such as most functional and
object-oriented languages, present the programmer with an abstract model
of data representations.
While such an abstraction hides the
details of the special run-time representations needed to support high-level
features, it comes at the cost of making interoperability with
low-level languages like \C{} non-trivial.
This incompatibility poses serious challenges for both implementors and
users of high-level languages, since there are numerous important libraries
that have \C{} APIs (application program interfaces).
For the purposes of this paper, we view \C{} as the prototypical
low-level language.\footnote{
  We do not consider the problems associated with interoperability
  between different high-level languages in this paper.
}

All widely used high-level language implementations provide some means
for calling {\em foreign functions} written in \C{}.  Such a mechanism
is called a {\em foreign-function interface} (FFI).  The requirements
of an FFI mechanism are to convert the arguments of a call from their
high-level to their low-level representations (called {\em
marshaling}), to handle the transfer of control from the high-level
language to \C{} and back, and then to convert the low-level
representation of the results into their corresponding high-level
representation (called {\em unmarshaling}).  In addition, the FFI
mechanism may map errors to high-level exceptions.  The details of how
data marshaling and unmarshaling are performed define a {\em policy}
that determines how high-level code can interact with foreign functions.

One important policy question is how to treat complicated foreign data
structures, such as \C{} \kw{structs}, arrays, and pointer data
structures.  While most existing FFI mechanisms handle \C{} scalars
well, they usually treat large \C{} data structures as abstract values
in the high-level language.  Although this approach is often
sufficient, there are situations in which the high-level language
needs to have direct access to large \C{} data structures and
marshaling is infeasible.
Examples include situations in which
an outside authority predetermines a data format not expressible in
the high-level language, such as network
packet headers and RPC stub generation, or where the volume of data
makes data marshaling prohibitively expensive, such as real-time
graphics applications that must pass large amounts of vertex and
texture data to the rendering engine.
For these situations, we need a
{\em foreign-data interface} (FDI), which is a mechanism for allowing
the high-level language to manipulate \C{} representations directly.
The combination of an FFI and an FDI provides a complete solution to the
interoperability problem.

This paper describes the architecture of the \moby{} compiler and how
that architecture efficiently supports a wide range of
interoperability policies, including both FFIs and FDIs.
\moby{} is a high-level,
statically-typed programming language with an \ml{}-like module
system~\cite{moby-classes}. \moby{}'s support for interoperability is
based on two features of its compiler.  First, the compiler's
intermediate representation, called BOL, is expressive enough to
implement \C{}.
Hence, it is easy for interoperability
tools to generate BOL code to manipulate \C{} data structures, access
\C{} global variables, and invoke \C{} functions.
Second, the compiler can import BOL code into its compilation environment;
including BOL code produced by interoperability tools.

This framework produces very efficient foreign interfaces because
the compiler infrastructure supports cross-module inlining of BOL
code, allowing it to tightly integrate BOL code for 
manipulating \C{} structures with
high-level client code.  Because of inlining,
the assembly code sequences generated by the \moby{} compiler
for accessing \C{} data structures have very little overhead;
indeed, the instruction sequences mimic those produced by a \C{}
compiler.  We use the same mechanism to define \moby{}'s primitive
types, such as integers, and the associated operations,
such as addition.  Hence the efficiency of this
mechanism is crucial to the performance of the \moby{} compiler.

Our architecture serves as the basis for implementing a number of
different foreign interface policies.  Each such policy must
determine how \C{} types should be packaged in \moby{}.  We have built
tools that 
implement two different policies: one that marshals data to cross the
boundary between \C{} and \moby{} and one that simply embeds \C{} into
\moby{}.
The first of these belongs to the family of
interoperability tools based on an \textit{interface description
language} or IDL~\cite{h-direct,caml-idl,ml-idl}. 
An IDL provides annotations to specify function argument and
result-passing conventions, as well as the semantics of \C{} types
(\eg{}, marking a ``\texttt{char} \texttt{*}'' value as a string).  
The \TOOL{moby-idl} tool generates a \moby{} interface to \C{}
functions from an IDL specification.    The second
tool, called \charon{}, implements the minimal (or identity)
policy.  By embedding \C{} directly into \moby{}, this tool
provides both a foreign function interface and a foreign data
interface, in that it allows high-level code to manipulate low-level
data structures in-place.  
It is worth noting that in our framework, a given \moby{} program may
use foreign APIs from both of these sources, or others like them,
concomitantly. 

\cut{
In summary, our approach relies on a low-level mechanism for data
interoperability, upon which we build tools that support various
policies for mapping from low-level APIs to \moby{}.  While supporting
data-level interoperability requires compiler support, our approach is
not specific to the \moby{} language {\em per se}.
}

The paper is organized as follows.
In \secref{sec:infrastructure}, we describe the architecture of the
\moby{} compiler and explain how this architecture supports data-level
interoperability.   
We describe \TOOL{moby-idl} and \charon{} in \secref{sec:tools} as
examples of interoperability tools built using our framework.
In \secref{sec:results},
we give experimental evidence showing that
these tools yield very efficient foreign interfaces between \C{} and
\moby{}. 
We discuss related work in \secref{sec:related} and conclude
in \secref{sec:concl}.

\newcommand{\word}{\textbf{word}}
\newcommand{\memory}{\textbf{memory}}
\newcommand{\var}{\textbf{var}}

\section{The \moby{} compiler infrastructure}
\label{sec:infrastructure}
\begin{figure}[t]
  \begin{center}
    \includegraphics[width=4in]{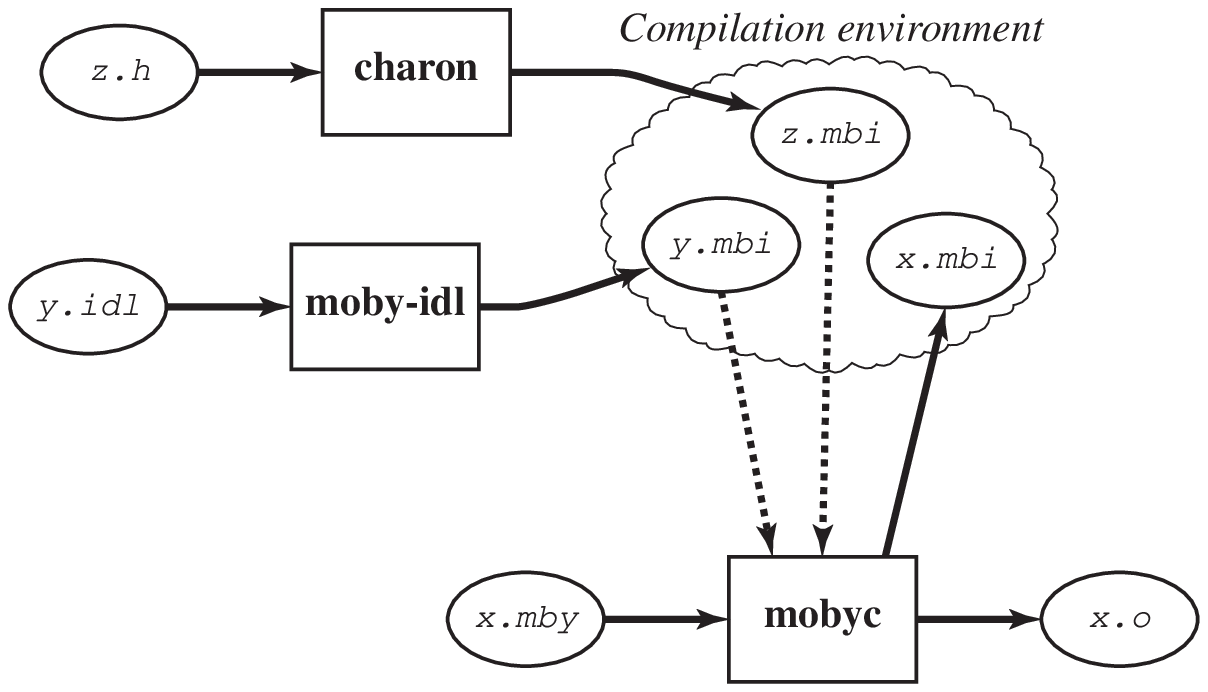}
  \end{center}
  \caption{The \moby{} compiler infrastructure}
  \label{fig:moby-toolchain}
\end{figure}

Our framework for interoperability is based on our existing compiler
infrastructure.
There are several aspects of this infrastructure that are key to supporting
interoperability:
\begin{itemize}
  \item
    The compiler's intermediate representation, called BOL, is
    expressive enough to describe low-level data representations and
    manipulations that are not expressible in \moby{} itself.
  \item
    Primitive \moby{} types and operations are defined in
    terms of BOL types and functions.
    These definitions are given in external \moby{} interface files,
    called MBI files, and play a \role{} similar to that of {\em native methods}
    in \java~\cite{java-native-interface} in that they allow \moby{} interfaces
    to be implemented by low-level code that cannot be written in \moby{}.
  \item
    The compiler can import and inline code from MBI files.
  \item
    There is a tool for generating MBI files from textual
    descriptions, called MBX files.
\end{itemize}
\figref{fig:moby-toolchain} illustrates our compiler infrastructure when
compiling a \moby{} source file (\ittt{x.mby}) that imports interfaces from
both an IDL file (\ittt{y.idl}) and a \C{} header file (\ittt{z.h}).
We use \mobyidl{} to generate \ittt{y.mbi} from the IDL file and
\charon{} to generate \ittt{z.mbi} from the \C{} header file.
In this section, we discuss this infrastructure and the small number
of additional features we added to support interoperability.

\subsection{MBI files}
In addition to an object file, the \moby{} compiler (\textbf{mobyc}) generates
an MBI file when compiling a source file.
Such a file contains information to support cross-module typechecking,
analysis, and inlining.
Collectively, the MBI files of an application are called its {\em compilation
environment}.
An MBI file contains the information found in a \moby{} signature, but it
also can contain information about the implementation that is not visible in the
signature, such as the representation of abstract types (as BOL types)
and the implementation of functions (as BOL terms).

While most MBI files are generated by the compiler from \moby{} source
files, the compiler can use MBI files from other sources.
For example, the primitive types and operations (\eg{}, \texttt{Int} and
\texttt{+}) are specified in hand-written MBI files,\footnote{
  Strictly speaking, we write an MBX file that is translated into
  a binary MBI file.
} which the compiler imports.
We use the same mechanism to import information
about foreign functions and data representations into the compiler.

\subsection{BOL}
The \moby{} compiler uses an extended $\lambda$-calculus, called BOL,
as its intermediate representation for optimization.
We first describe BOL's type system and how to connect \moby{} types to their
underlying BOL representation.
We then give an overview of BOL terms
and how they support efficient interoperability.

BOL has a weak type system that does not guarantee type safety
(one can think of BOL's type system as roughly equivalent to \C{}'s
without the recursive types).
The \moby{} compiler uses BOL types to guide the mapping of BOL
variables to machine registers, to provide representation information
for the garbage collector (we are using the Smith-Morrisett
{\em mostly-copying collector}~\cite{mostly-copying-collector}),
and to provide some sanity checking for the optimizer's
transformations.
For example, using BOL type constructors \kw{enum},
\kw{ptr}, \kw{vector}, and \kw{struct}, we can
define names for BOL types:
\begin{centercode}
\kw{typedef} char = \kw{enum}(0,255)
\kw{typedef} string_data = \kw{ptr}(\kw{vector}(1, char))
\kw{typedef} string = \kw{ptr}(\kw{struct} 8:4 (0: \kw{int}, 4: string_data))
\end{centercode}
A \texttt{char} is 
a value between 0 and 255, a \texttt{string\char`\_data} is a pointer to
a vector of unknown length storing \texttt{char}s (each of which has a
size of one byte), and a \texttt{string} is a pointer 
to a heap object consisting of an integer (the length) and a
\texttt{string\char`\_data} pointer.
To promote interoperability with \C{}, the runtime representation of string
data is null-terminated. We use BOL \kw{struct} types to describe heap
objects.  In the example,
the ``\texttt{8:4}'' notation on the BOL \kw{struct} type gives its size
and alignment, respectively; the
``\texttt{0:}'' and ``\texttt{4:}'' are the offsets of its fields.
These definitions assume a 32-bit architecture.

The MBI file format allows an abstract \moby{} type to be defined in terms
of a BOL type definition.
For example, the MBI file that implements the primitive \moby{} types
defines the \moby{} \texttt{String} type in terms of the primitive BOL type
\texttt{string} given above.
\begin{centercode}
\kw{type} String = \kw{prim} string
\end{centercode}

BOL types are stratified into three levels (or kinds): types with kind
\word{} describe those values that can be represented by a
general-purpose register on the target machine (\eg{}, scalars and
pointers); types with kind \var{} describe those values that can be
bound to a BOL variable, which includes the \word{} types; and types
with kind \memory{} describe values that can be held in memory, which
includes the \word{} and \var{} types.  Target-specific aspects of the
type system are captured by the kinding judgements.  For example, on a
32-bit architecture the 64-bit integer type has \var{} kind, whereas
on a 64-bit architecture, the 64-bit integer type has \word{} kind.

Syntactically, BOL is a direct-sytle extended $\lambda$ calculus.\footnote{%
  Strictly speaking, BOL is a normalized representation
  that requires all intermediate results (including literals) to be bound
  to fresh variables, but our tools for compiling MBI files from their
  textual description accepts more complicated subexpressions and performs
  the normalization by introducing new temporary variables.
  In this paper, we use the unnormalized representation, since it is easier to
  read.
}
It includes a primitive notion of heap object and primitive
operations for manipulating raw machine data types such as addresses
and 32-bit integers.
The binding form
\begin{centercode}
\cdmath{}\kw{let} \ittt{obj} : \ittt{ty} = \kw{alloc} ($\text{\ittt{x}}_1$, ..., $\text{\ittt{x}}_n$)
  ...
\end{centercode}
allocates a heap object of $n$ fields and binds \ittt{obj} to it.
The memory layout of \ittt{obj} is inferred from the types of
the $\text{\ittt{x}}_i$.
We use the notation \kw{\#}\ittt{i} to select the \ittt{i}th field
of such objects (zero-based indexing).
BOL primitive operations include \texttt{I32Add} for adding 32-bit
integers, \texttt{AdrAdd} for pointer arithmetic, \texttt{AdrEq} for
comparing addresses, and \texttt{AdrLoadI32} for fetching 32-bit integers
from memory.

MBI files can include definitions of \moby{} functions in terms of
BOL terms; these definitions are used to support cross-module inlining
and inlining of primitive operations.
For example, the \texttt{length} function on strings is defined
as follows:
\begin{centercode}
\kw{val} length : String \kw{->} Int =
      \kw{fun} len (s : string, _ : exn_handler) \{
        \kw{let} n : \kw{int} = s#0
        \kw{return} n
      \}
\end{centercode}
The first line gives the \moby{} name and type of the function;
the remaining code is its BOL definition. The BOL function
\texttt{len} (the name is to allow recursion) has two
parameters.  The first parameter (\texttt{s}) has BOL \texttt{string} type
(which is the representation of the \moby{} \texttt{String} type);
the second is the implicit exception-handler continuation 
parameter required by the
\moby{} calling convention.
Since the implementation of the \texttt{length} function does not use
this parameter, we use a wildcard for it.
The body of the function is trivial: we select the first component of
\texttt{s} (recall that the BOL type string is a pointer to a heap
object), bind it to \texttt{n}, and then return \texttt{n}.

The \moby{} compiler inlines such BOL implementations
to reduce the overhead of calling the associated \moby{} functions
and to enable other optimizations.
For example, when compiling the \moby{} expression
\texttt{length s + 1},
the definition of \texttt{length} will be inlined, resulting in
the following intermediate BOL term:
\begin{centercode}
\kw{let} n = s#0
\kw{let} t = I32Add(n, 1)
\kw{return} t
\end{centercode}
In the case that \texttt{s} is bound to a known literal, this expression
can be further reduced to a constant.

\subsection{Interoperability extensions}
To support interoperability, we have added \C{} calls, function types,
and declarations to BOL. To  improve the efficiency of foreign
function calls, we added a mechanism to stack allocate temporary
storage.  We describe these pieces in turn.

BOL has a binding form for calling \C{} functions:
\begin{centercode}
\cdmath{}\kw{let} \ittt{result} = \kw{ccall} \ittt{id} ($\text{\ittt{arg}}_1$, ..., $\text{\ittt{arg}}_n$)
  ...
\end{centercode}%
The identifier \ittt{id} is a BOL variable that may be bound to a
known \C{} function (or to a function pointer).
The type of \texttt{id} controls the calling convention for the \C{} call;
we added a \C{}-function type constructor to the BOL type system
for this purpose.
The types of \C{} function parameters are mapped to BOL types in a way
that preserves the specified calling convention (\eg{}, small integer
types are promoted to the BOL equivalent of \kw{int}).
The one technical complication arises from \kw{struct} parameters.
BOL's kinding system does not allow BOL variables to be bound to values whose
type has \memory{} kind, so we have to represent \C{} \kw{struct}
values by their addresses.
The problem then is how to distinguish between a pointer to a \kw{struct} and
a \kw{struct} parameter?
We added an additional type constructor for the latter case to solve
this problem.
We also added a \kw{void} type to represent \kw{void} return types
from \C{} functions.
Our mechanism does not support \emph{varargs} yet, but it should be possible
to do so once the underlying code generator has such support.\footnote{
  Our code generator is based on the MLRisc framework~\cite{mlrisc}.
}

If an MBI file has a reference to a \C{} function, it must contain
an external declaration, which has the following form in the textual (MBX)
representation: 
\begin{centercode}
\cdmath\kw{external} \ittt{result-ty} \textit{C-ident} ($\text{\ittt{param-ty}}_1$, ..., $\text{\ittt{param-ty}}_n$)
\end{centercode}
where \ittt{result-ty}, $\text{\ittt{param-ty}}_1$, ..., $\text{\ittt{param-ty}}_n$
are BOL types of \var{} kind.

Connecting a high-level language and \C{} often requires temporary
space for marshaling \kw{struct} arguments and results.
Efficiency considerations require a lightweight mechanism for
allocating this space. 
Since the lifetime of such storage is the call to the \C{} function,
the most efficient place for such storage is in the stack frame of the
wrapper function. 
To support such allocation, we added a binding form to BOL for
allocating stack space.
The binding form
\begin{centercode}
\cdmath\kw{stackalloc} \textit{x}[$\mathit{sz}$, $\mathit{align}$]
  ...
\end{centercode}
binds \ittt{x} to $\mathit{sz}$ bytes of stack
storage aligned on an $\mathit{align}$-byte
boundary ($\mathit{sz}$ and $\mathit{align}$ are both integer constants).
The lifetime of the storage is the scope of the binding
(\ie{}, the ``\texttt{...}''), so BOL code that uses \kw{stackalloc}
must not allow references to the storage to escape.
In practice, this restriction has not been a problem; furthermore, the
\moby{} optimizer does not perform any transformations that would expand
the extent of a variable beyond its scope.

\section{Applications}
\label{sec:tools}
Our framework is  designed to support a wide range of
interoperability policies.
In this section, we describe two foreign-interface generator
tools for \moby{} that implement significantly different policies
on top of our framework.
We also describe some other potential uses of our framework.

\subsection{Moby IDL}
\label{sec:tools-idl}

One of the most popular ways of automating the connection between
high-level languages and \C{} is to use an \emph{interface description language}
(IDL) to specify the foreign interface~\cite{h-direct,caml-idl,ml-idl}.
An IDL specification is essentially a \C{} header file with annotations.
The annotations are used to specify the {\em direction} of parameters
and the interpretation of pointer types.
IDL-based generators compile these specifications into high-level
interfaces and glue code.
We have retargeted the \smlnj{} IDL-based foreign-interface
generator~\cite{ml-idl} to produce MBI files from an
IDL specification.\footnote{
  There are a number of IDL variants; the \smlnj{} tool (and \TOOL{moby-idl})
  accepts an extension of the OSF DCE dialect, which is essentially
  the version that Microsoft uses for COM~\cite{essential-idl}.
}

The \TOOL{moby-idl} tool generates an MBI file that defines a
\moby{} type for each non-trivial type in the IDL
specification and a stub function for each function prototype.
The stub function embodies a \emph{copy-in/copy-out} policy for calling \C{}
functions: every input parameter is translated from its \moby{} representation
to \C{} and then passed to the \C{} function, and the result and every
output parameter is translated from \C{} to its \moby{} representation.
The stub functions have \moby{} types, but are implemented using BOL.

To make this discussion concrete, we give two examples, each
highlighting different aspects of the \mobyidl{} tool.
The first example is the \texttt{getenv} function from the \C{} library.
This function takes a string argument that names an environment variable
and returns the value of the variable as a string.
If the named variable has no value, \texttt{NULL} is returned.
We use the following IDL specification to capture this behavior:
\begin{centercode}
\kw{typedef} [unique,string] char *StringOpt;

StringOpt getenv ([in,string] char *name);
\end{centercode}
The \texttt{StringOpt} type is annotated as being \texttt{unique}, which
means that it can be \texttt{NULL}, and it is annotated as being a \texttt{string}.
The \texttt{name} parameter to the \texttt{getenv} function is marked as being
an input parameter and also as a string.
The \mobyidl{} tool is guided by these annotations to produce the
following \moby{} specification:
\begin{centercode}
\kw{val} getenv : String \kw{->} Option(String)
\end{centercode}
In the MBI file, this stub function has the following definition:
\begin{centercode}
\kw{val} getenv : String \kw{->} Option(String) =
      \kw{fun} getenv (str : string, _ : exn_handler) \{
	\kw{let} c_str = str#1
	\kw{let} c_res = \kw{ccall} getenv (c_str)
	\kw{if} AdrEq (c_res, \kw{nil}) \kw{then} \kw{return} 0
	\kw{else} \{
          \kw{let} res_str = \kw{ccall} MOBY_AllocCString(c_res)
	  \kw{let} res = \kw{alloc} (res_str)
	  \kw{return} (res)
	\}
      \}
\end{centercode}
This BOL code converts the input parameter \texttt{str} to a \C{} string
by selecting the second component of the heap object (recall that
\moby{} strings are represented as a pair of a length and data-pointer)
and passing it to the \texttt{getenv} function.
If the result (\texttt{c\char`\_res}) is \texttt{NULL}, then
\texttt{0} is returned, which is the representation of the \moby{}
constant \texttt{None}.
If the result is not \texttt{NULL}, then a \moby{} string is allocated
and initialized from \texttt{c\char`\_res} by calling the \moby{}
runtime system function \texttt{MOBY\char`_AllocCString}.
The \moby{} string is then wrapped in a one-word heap object (the
representation of the \texttt{Some} data constructor) and returned.

A slightly more involved example is the following IDL
specification of the interface to the \textsc{Unix}
\texttt{gettimeofday} system call: 
\begin{centercode}
\kw{typedef} \kw{struct} \{
    long  tv_sec;
    long  tv_usec;
\} timeval;

\kw{typedef} \kw{struct} \{
    int  tz_minuteswest;
    int  tz_dsttime;
\} timezone;

int gettimeofday (
    [ref, out] timeval *t,
    [ref, out] timezone *tz);
\end{centercode}
This specification includes \texttt{out} annotations on the two parameters
of \texttt{gettimeofday}.
These \texttt{out} annotations identify the parameters as pointers to
storage for returning the results of the function call.
The \TOOL{moby-idl} tool produces the following \moby{} interface
from this specification:
\begin{centercode}
\kw{datatype} Timeval \{ TIMEVAL \kw{of} (Int, Int) \}
\kw{datatype} Timezone \{ TIMEZONE \kw{of} (Int, Int) \}
\kw{val} gettimeofday : () \kw{->} (Int, Timeval, Timezone)
\end{centercode}
The tool has translated the \C{} \kw{struct} types to \moby{}
singleton datatypes.\footnote{
  A labeled record type would be preferable, but \moby{}
  does not have records yet.
}
The tool has mapped the \texttt{out} parameters 
to results in generating the \moby{} type  for the function
\texttt{gettimeofday}. 
Again, we implement the stub code that connects \C{} and \moby{} 
using BOL code in the generated MBI file. 
This code has the following form:
\begin{centercode}
\kw{val} gettimeofday : () \kw{->} (Int, Timeval, Timezone) =
      \kw{fun} gettimeofday (_ : exn_handler) \{
	\kw{stackalloc} tm[8:4], tz[8:4]
	\kw{let} res = \kw{ccall} gettimeofday (tm, tz)
	\kw{let} tm2 = \kw{alloc}(AdrLoadI32(tm), AdrLoadI32(AdrAdd(tm, 4)))
	\kw{let} tz2 = \kw{alloc}(AdrLoadI32(tz), AdrLoadI32(AdrAdd(tz, 4)))
	\kw{return} (res, tm2, tz2)
      \}
\end{centercode}
The code uses the BOL \kw{stackalloc} construct to allocate
temporary storage for the results in the stack.
The variables \texttt{tm} and \texttt{tz} are each bound to
the address of 8~bytes of stack storage (with 4-byte alignment).
The extent of this storage is the scope of the variables (the rest of
the function in this case).
After calling the \C{} \texttt{gettimeofday} function, the code
extracts the contents of the results from the temporary storage and allocates
a pair of heap objects for the results.

In our framework, uses of the \texttt{getenv} or \texttt{gettimeofday}
functions can be inlined at the call site, which avoids the extra level
of function call found in most IDL-based FFI mechanisms and also
provides further opportunities for optimization such as avoiding unnecessary
marshaling.
\secref{sec:results} provides experimental results that demonstrate the
efficiency of this approach.

\subsection{Charon}
\label{sec:tools.charon}
Using the Moby compiler infrastructure, we have built a second
interoperability tool, called \charon{}.  This tool implements the
minimal interoperability policy, in that it simply embeds \C{} into
\moby{}.  It maps \C{} types into abstract \moby{} types and provides
\moby{} functions, implemented in \bol{}, for manipulating \C{}
values.  It provides \moby{} functions, again implemented in \bol{},
for calling \C{} functions.

\charon{} takes as input a \C{} header file and produces two output
files.  The first file contains a \moby{} signature describing the
types and operations defined by the header file.  The second file is
an \mbi{} file containing \bol{} code that implements the signature in
the first file.  Note that the signature cannot be implemented in
\moby{} directly because unlike \bol{}, \moby{} does not include the
low-level operations necessary to manipulate \C{} data structures.
The \moby{} compiler converts \mbi{} files into assembly code that can
be linked with the compiled \C{} code that implements the header file.  

\charon{} factors its embedding into two parts: one generic to \C{}
and one specific to the input header file.
The generic part is the \emph{C-interface library}, which is implemented
by a hand-written MBI file.
The \C{}-interface library provides several type constructors for
encoding \C{} types:
\begin{centercode}
  \kw{type} LValue(t)
  \kw{type} CPtr(t)
  \kw{type} SizeOf(t)
\end{centercode}
\texttt{LValue(ty)} is the type of an assignable \C{}
value of type \texttt{ty}.   
The underlying representation of \texttt{LValue(ty)} is an address of
a memory-location containing a value of type \texttt{ty} (\eg{},
the address of a \C{} global variable or a \C{} heap location).
\texttt{CPtr(ty)} is the type of a
pointer to a value (or array of values) of type \texttt{ty}.
\texttt{SizeOf(ty)} is used to type an abstract
representation of the size of \texttt{ty}; we explain further below.
We define a collection of \emph{phantom} types\footnote{
  Phantom types are types whose only purpose is to serve as arguments
  to type constructors~\cite{Burton90,reppy:safe-sockets,Leijen99}.
  The idea of using phantom types to encode \C{} types was suggested
  to us by Matthias Blume~\cite{blume:c-in-ml}.
}
corresponding to the primitive \C{} types (\eg{}, \texttt{SChar} for signed
characters and \texttt{SInt} for signed integers).
These phantom types are used to constrain the types of generic operations
on \C{} values.
For example, the \C{}-interface library defines the following
operations on \C{} pointers:
\begin{centercode}
  \kw{val} isNull : [t] CPtr(t) \kw{->} Bool 
  \kw{val} deref  : [t] CPtr(t) \kw{->} LValue(t) 
  \kw{val} getPtr : [t] LValue(CPtr(t)) \kw{->} CPtr(t) 
  \kw{val} setPtr : [t] (LValue(CPtr(t)), CPtr(t)) \kw{->} ()
  \kw{val} malloc : [t] SizeOf(t) \kw{->} CPtr(t)
\end{centercode}
(the notation ``\texttt{[t]}'' is the binding of a type variable
\texttt{t}).
The \C{}-interface library also defines specific operations
for the primitive \C{} types, such as the following operations
for signed integers:
\begin{centercode}
  \kw{val} getSInt : LValue(SInt) \kw{->} Int 
  \kw{val} setSInt : (LValue(SInt), Int) \kw{->} () 
  \kw{val} sizeOfSInt : () \kw{->} SizeOf(SInt)
\end{centercode}

In addition to the generic support provided by the \C{}-interface library,
\charon{} generates an MBI file that contains types and functions
specific to the input header file.
For example, consider the following \C{} declarations that
describe a binary tree type and a function for creating such trees: 
\begin{centercode}
\kw{typedef} \kw{struct} tree \{
    int         label;
    tree_ptr    left;
    tree_ptr    right;
\} tree_node, *tree_ptr;
\kw{extern} tree_ptr MakeTree (int depth);
\end{centercode}%
From these declarations, \charon{} generates an MBI file
that implements the following \moby{} interface:
\begin{centercode}
\kw{type} Struct_tree
\kw{type} Def_tree_node = Struct_tree
\kw{type} Def_tree_ptr = CPtr(Struct_tree)

\kw{module} Stree \{
  \kw{val} label  : LValue(Struct_tree) \kw{->} LValue(SInt) 
  \kw{val} left   : LValue(Struct_tree) \kw{->} LValue(Def_tree_ptr) 
  \kw{val} right  : LValue(Struct_tree) \kw{->} LValue(Def_tree_ptr) 
  \kw{val} sizeOf : () \kw{->} SizeOf(Struct_tree) 
\}

\kw{val} makeTree : Int \kw{->} LValue(Def_tree_ptr) 
\end{centercode}
The type \texttt{Struct\char`\_tree} is a phantom type corresponding
to the \C{} \kw{struct} \texttt{tree} type.
The two \C{} type definitions (\texttt{tree\char`\_node} and
\texttt{tree\char`\_ptr}) are translated to \moby{} type definitions.
The \kw{struct} \texttt{tree} type is mapped to a module of operations
for accessing its fields.
Note that the access functions return \texttt{LValue}s, which can be used
to assign to the fields.
In addition, a \texttt{sizeOf} function is provided, which can be used
to allocate objects of  type  \texttt{Struct\char`\_tree}.
Lastly, the \texttt{MakeTree} function is mapped to the \moby{} function
\texttt{makeTree}.

Using the C-interface library and the generated interface, we can
write \moby{} code that manipulates the \texttt{tree} data structures
and calls the \texttt{MakeTree} function.
For example, the following \moby{} function walks a tree, adding one to each
label:
\begin{centercode}
\kw{fun} incLabels (t : CPtr(Struct_tree)) \kw{->} ()
\{
  \kw{if} isNull t
    \kw{then} ()
    \kw{else} \{
      \kw{val} t = deref t;
      setSInt(Stree.label t, getSInt(Stree.label t) + 1);
      incLabels (getPtr(Stree.left t));
      incLabels (getPtr(Stree.right t))
    \}
\}
\end{centercode}

The generated \moby{} interface is implemented using BOL types and functions
in the generated MBI file.
For this interface, the MBI file contains the following definitions:
\begin{centercode}
\kw{type} Struct_tree = \kw{prim} \kw{void}
\kw{type} Def_tree_node = Struct_tree
\kw{typedef} def_tree_ptr = \kw{addr}(\kw{data})
\kw{type} Def_tree_ptr = \kw{prim} def_tree_ptr
\end{centercode}
Notice that the phantom type \texttt{Struct\char`\_tree} is
defined to be the BOL \kw{void} type.
Because BOL does not have recursive types, we map the \C{} \texttt{tree\char`\_ptr}
type to \texttt{\kw{addr}(\kw{data})} (the BOL equivalent of \texttt{\kw{void}*}).
The access functions for the \kw{struct} \texttt{tree} type
are implemented as follows:
\begin{centercode}
\kw{module} Stree \{
  \kw{val} label : LValue(Struct_tree) -> LValue(SInt) =
        \kw{fun} fld(p : lvalue, _ : exn_handler) \{ \kw{return} p \}

  \kw{val} left : LValue(Struct_tree) -> LValue(Def_tree_ptr) =
        \kw{fun} fld(p : lvalue, _ : exn_handler)
        \{ \kw{let} q = AdrAdd(p, 4) \kw{return} q \}
    
  \kw{val} right : LValue(Struct_tree) -> LValue(Def_tree_ptr) =
        \kw{fun} fld(p : lvalue, _ : exn_handler)
        \{ \kw{let} q = AdrAdd(p, 8) \kw{return} q \}

  \kw{val} sizeOf : () -> SizeOf(Struct_tree) =
        \kw{fun} sz(_ : exn_handler) \{ \kw{let} n = 12 \kw{return} n \}
\}
\end{centercode}
Finally, the \texttt{makeTree} function is implemented as a trivial
wrapper around the \texttt{MakeTree} \C{} function.
\begin{centercode}
\kw{extern} \kw{addr}(struct_tree) MakeTree (\kw{int})
\kw{val} makeTree : Int -> CPtr(Struct_tree) =
      \kw{fun} makeTree (arg : \kw{int}, _ : exn_handler) \{
        \kw{let} result : \kw{addr}(struct_tree) = \kw{ccall} MakeTree(arg)
        \kw{return} result
      \}
\end{centercode}

\subsection{Other applications}
Other approaches to generating foreign interfaces are also compatible
with our framework.  For example, the \TOOL{C${\rightarrow}$Haskell}
tool uses an interface specification file coupled with a \C{} header
file to generate a Haskell interface to a \C{}
library~\cite{c->haskell}.  The Haskell interface is implemented using
GHC's foreign interface support.  Our framework can express the same
level of interoperability, so building a \TOOL{C${\rightarrow}$Moby}
tool should be straightforward.

A more interesting application of our framework is to support
domain-specific data objects.
For example, a tool for generating evaluators from
atttribute-grammar specifications might use a
highly-tuned tree representation that cannot be expressed
directly in \moby{}.
Such a tool could instead generate an MBI file that defines
its tree representation, along with operations on the trees,
which can then be used by other code written in \moby{}.
The compiler's cross-module inlining mechanism can then
eliminate any performance penalty for using an abstract type.

\section{Experimental results}
\label{sec:results}

In this section, we present the results of some synthetic benchmarks
that demonstrate the efficiency of interoperability in our framework.
Our measurements were performed on a dual-733MHz PIII workstation running
Linux, kernel version 2.2.14.

Our first benchmark tests the overhead of marshaling when using the
\mobyidl{} tool.
For this benchmark, we measured the time taken to perform $10^7$ calls
of the \texttt{gettimeofday} system call.
The interface to \texttt{gettimeofday} was generated from
the IDL specification given as an example in \secref{sec:tools-idl}.
We compare the performance of the \mobyidl{} tool with that of
the Haskell \TOOL{H/Direct} tool (using GHC 4.08.2)
and \TOOL{camlidl} (using OCAML 3.00),
as well as with the direct \C{} version (using gcc -O2, version 2.91.66).
The measured user and system times for this benchmark are given in
\tblref{tbl:tod}.
\begin{table}[tp]
  \caption{Measured execution times for \texttt{gettimeofday} benchmark}
  \label{tbl:tod}
  \begin{center}\small
    \begin{tabular}{|l|ccc|cc|}
    \hline
      Language &
      \multicolumn{3}{|c}{Execution time (seconds)} &
      \multicolumn{2}{|c|}{Relative to C}
    \\
      or tool &
      sys & usr & tot &
      usr & tot
    \\\hline
      C		& 5.02 & 1.96 & 6.98 & 1.00 & 1.00 \\
      moby-idl	& 5.18 & 2.25 & 7.43 & 1.15 & 1.06 \\
      camlidl	& 5.48 & 4.90 & 10.38 & 2.50 & 1.48 \\
      H/Direct	& 6.40 & 68.68 & 75.08 & 35.0 & 10.76 \\
    \hline
    \end{tabular}
  \end{center}
\end{table}%
The first three data columns give the execution time in seconds (user, system,
and total).
The last two columns give the ratio of the user and total times compared to
the direct \C{} version.
As expected, each version of the test uses essentially the same amount
of system time (about 5 seconds), but they have very different user times.
Discounting the system time, the \mobyidl{} version has a marshaling
overhead of about 15\% over the direct \C{} version.
This result compares favorably with the \TOOL{camlidl} overhead of 250\% and
the \TOOL{H/Direct} overhead of 3500\%.\footnote{
  We believe that the high execution time of the \TOOL{H/Direct} version is
  caused by its use of \texttt{malloc} and \texttt{free} to manage
  the temporary storage needed for the results of the \texttt{gettimeofday}
  call.
}

Our second benchmark is designed to test the efficiency of direct
access to \C{} data structures from \moby{} code using our framework.
Each iteration of the benchmark constructs a complete binary tree
of depth 16 (65535 nodes), where each node is
labeled with an integer generated by calling the \texttt{rand}
function from the \C{}-library.
After constructing the tree, we perform a depth-first traversal
to find the largest label, and then explicitly free the tree.
The benchmark iterates these three steps 100 times.
We wrote the \moby{} version of the benchmark entirely in \moby{}
using the \charon{}-generated interface to the \C{} representation.
The tree data structure was managed using the \moby{} C-interface
library's interface to \texttt{malloc} and \texttt{free}.
We compare the performance of the \moby{} program with the native \C{}
version of the same algorithm; the results are given in \tblref{tbl:tree}.
\begin{table}[tp]
  \caption{Measured execution times for \texttt{tree} benchmark}
  \label{tbl:tree}
  \begin{center}\small
    \begin{tabular}{|l|ccc|cc|}
    \hline
      Language &
      \multicolumn{3}{|c}{Execution time (seconds)} &
      \multicolumn{2}{|c|}{Relative to C}
    \\
      or tool &
      sys & usr & tot &
      usr & tot
    \\\hline
      C		& 0.37 & 4.61 & 4.98 & 1.00 & 1.00 \\
      charon	& 0.29 & 4.36 & 4.65 & 0.94 & 0.93 \\
      GHC	& 0.41 & 15.10 & 15.51 & 3.28 & 3.11 \\
    \hline
    \end{tabular}
  \end{center}
\end{table}%
From these results, we can see that the data-level interoperability
supported by \charon{} has no overhead over native \C{} code.\footnote{
  The slight performance advantage that \moby{} has in this benchmark
  is most likely a result of a slightly more efficient argument passing mechanism.
}
We also measured a version of the program compiled under GHC.
In this version, we used GHC's \texttt{Addr} type and strictness
annotations to implement
an interface similar to the one provided by \charon{}.
These measurements demonstrate that while GHC's foreign-data
support has much of the expressiveness of our framework, it lags in
efficiency.

\section{Related work}
\label{sec:related}

Our approach to interoperability is based on the \moby{} compiler
infrastructure.
This infrastructure serves as the foundation for a wide range of
interoperability policies, each with its own user-level
mechanism.
This approach contrasts with most of the prior work on
language interoperability, which fixes a particular interoperability
policy and user-level mechanism.\footnote{%
  Of course, some systems have multiple mechanisms,
  each of which supports a different policy.
}

Most high-level language implementations provide some mechanism for
connecting with \C{} code.  Often, this mechanism requires
hand-written stub functions to translate between the
high-level and \C{} representations.
Examples of languages with such mechanisms 
include \java{} (the \java{} Native
Interface)~\cite{java-native-interface}, \smlnj{}, and 
\ocaml{}~\cite{ocaml-manual-3.0}.
Our framework also supports hand-written stubs.
Such stubs can either be written in \C{}, as we have done with
some low-level, run-time system functions, or in an MBX file.
In the latter case, the \moby{} compiler's cross-module inlining mechanism
allows the stub code to be inlined at its call sites.

Some systems make it possible to write interoperability code in the
high-level language.
For example, both the \sml{}'97 Basis Library~\cite{sml-basis-library} and
the Glasgow \haskell{} compiler (GHC)~\cite{ghc-manual} provide
operations for reading and writing scalar values in a \emph{bytearray}.
With this mechanism, 
one can manipulate a \C{} data structure by importing it into a program
as a bytearray.
However, the user is responsible for understanding the layout
of the data structure.
The GHC mechanism is lower-level than that of the \sml{} Basis;
specifically, GHC does no bounds checking and it is possible to read and
write pointer values.

In many respects, GHC's foreign interface support is the closest to
providing the flexibility of our framework.
Our mechanism is slightly more expressive in that we support \C{} functions
with \kw{struct} arguments.
As demonstrated in \secref{sec:results}, we also have a significant
performance advantage.
For more complicated policies, such as those required by IDL,
the BOL \kw{stackalloc} construct can greatly reduce
the overhead of data marshaling.
It would difficult to put such a mechanism into a high-level
language without using a sophisticated type system to control the
extent of stack-allocated variables.

\section{Conclusion}
\label{sec:concl}
In this paper, we described \moby{}'s interoperability infrastructure,
which consists of an expressive intermediate representation called BOL
and the ability to import externally-defined BOL code into the
compilation environment.
This infrastructure supports a wide-range of
interoperability tools, from IDL-based tools that marshal their data
to tools such as \charon{} that provide data-level interoperability by
embedding \C{} into \moby{}.  This infrastructure is highly
efficient because it supports cross-module inlining of BOL code.
Experimental evaluation showed that the overhead associated with
invoking foreign functions and manipulating \C{} data structures was
very low.
While supporting data-level interoperability requires compiler support,
our approach is not specific to the \moby{} language {\em per se}.
Specifically, we expect that our approach is compatible with compilers
that have more strongly-typed IRs.

\bibliographystyle{small-alpha}
\bibliography{refs}

\end{document}